\documentclass[conference]{IEEEtran}
\IEEEoverridecommandlockouts
\usepackage[utf8]{inputenc} 
\usepackage[T1]{fontenc}
\usepackage{url}
\usepackage{ifthen}
\usepackage{cite}
\usepackage[cmex10]{amsmath} %

\usepackage[acronym]{glossaries}
\newacronym{acf}{ACF}{autocorrelation function}
\newacronym{air}{AIR}{achievable information rate}
\newacronym{app}{APP}{a posteriori probability}
\newacronym{ase}{ASE}{amplified spontaneous emission}
\newacronym{ask}{ASK}{amplitude-shift keying}
\newacronym{awgn}{AWGN}{additive white Gaussian noise}

\newacronym{b2b}{B2B}{back-to-back}
\newacronym{baa}{BAA}{Blahut-Arimoto Algorithm}
\newacronym{bpcu}{bpcu}{bits per channel use}

\newacronym{cd}{CD}{chromatic dispersion}
\newacronym{cdc}{CDC}{chromatic dispersion compensation}
\newacronym{cdf}{cdf}{cumulative density function}
\newacronym{clt}{CLT}{central limit theorem}
\newacronym{coi}{COI}{channel of interest}
\newacronym{cpan}{CPAN}{correlated phase and additive noise}
\newacronym{cspr}{CSPR}{carrier to signal power ratio}
\newacronym{cscg}{CSCG}{circularly symmetric complex Gaussian}
\newacronym{csi}{CSI}{channel state information}
\newacronym{cwt}{CWT}{continouus-wave tone}

\newacronym{dbp}{DBP}{digital backpropagation}
\newacronym{dsp}{DSP}{digital signal processing}
\newacronym{dtft}{DTFT}{discrete time Fourier transform}

\newacronym{dd}{DD}{direct detection}
\newacronym{dft}{DFT}{discrete Fourier transform}
\newacronym{dmd}{DMD}{differential mode delay}
\newacronym{dmgd}{DMGD}{differential mode group delay}
\newacronym{dgd}{DGD}{differential group delay}
\newacronym{dt}{DT}{decision tree}

\newacronym{eepn}{EEPN}{equalization-enhanced phase noise}
\newacronym{ep}{EP}{expectation propagation}
\newacronym{epi}{EPI}{entropy power inequality}
\newacronym{erp}{ERP}{enhanced regular perturbation}
\newacronym{evd}{EVD}{Eigenvalue decomposition}

\newacronym{fer}{FER}{frame error rate}
\newacronym{ffe}{FFE}{feedforward equalizer}
\newacronym{fft}{FFT}{fast Fourier transform}
\newacronym{fmf}{FMF}{few-mode fiber}
\newacronym{fwhm}{FWHM}{full-width half-maximum}
\newacronym{fwm}{FWM}{four-wave mixing}

\newacronym{ghq}{GHQ}{Gauss-Hermite-Quadrature}
\newacronym{gmi}{GMI}{generalized mutual information}
\newacronym{gvamp}{GVAMP}{generalized \gls{vamp}}

\newacronym{ici}{ICI}{intercarrier interference}
\newacronym{id}{ID}{informational divergence}
\newacronym{idra}{IDRA}{ideally distributed Raman amplification}
\newacronym{iid}{i.i.d.}{independent and identically distributed}
\newacronym{im}{IM}{intensity modulation}
\newacronym{isi}{ISI}{intersymbol interference}
\newacronym{iss}{ISS}{intersymbol sample}

\newacronym{jd}{JD}{joint detection}
\newacronym{jdd}{JDD}{joint detection and decoding}
\newacronym{jdabaa}{DABAA}{dynamic-assignment Blahut-Arimoto Algorithm}

\newacronym{kk}{KK}{Kramers-Kronig}

\newacronym{lhs}{l.h.s.}{left-hand side}
\newacronym{lmmse}{LMMSE}{linear minimum-mean-square-error}
\newacronym{ls}{LS}{least squares}
\newacronym{lstm}{LSTM}{long short-term memory}
\newacronym{lp}{LP}{linearly polarized}
\newacronym{lut}{LUT}{look-up table}

\newacronym{map}{MAP}{maximum a-posteriori}
\newacronym{mcf}{MCF}{multi-core fiber}
\newacronym{md}{MD}{modal dispersion}
\newacronym{mgdm}{MGDM}{mode-group-division multiplexing}
\newacronym{mdg}{MDG}{mode-dependent gain}
\newacronym{mdl}{MDL}{mode-dependent loss}
\newacronym{mdm}{MDM}{mode-division multiplexing}
\newacronym{me}{ME}{Manakov equation}
\newacronym{mi}{MI}{mutual information}
\newacronym{mir}{MIR}{mutual information rate}
\newacronym{mimo}{MIMO}{multiple-in multiple-out}
\newacronym{miso}{MISO}{multiple-in single-out}
\newacronym{ml}{ML}{machine learning}
\newacronym{mmf}{MMF}{multi-mode fiber}
\newacronym{mmse}{MMSE}{minimum-mean-squared-error}
\newacronym{mpc}{MPC}{minimum-phase criterion}
\newacronym{mse}{MSE}{mean-squared error}
\newacronym{mzm}{MZM}{Mach-Zehnder modulator}

\newacronym{nlse}{NLSE}{nonlinear Schrödinger equation}
\newacronym{nmse}{NMSE}{normalized MSE}
\newacronym{nn}{NN}{neural network}

\newacronym{ofdm}{OFDM}{orthogonal frequency-division multiplexing}

\newacronym{pam}{PAM}{pulse-amplitude modulation}
\newacronym{pdf}{pdf}{probability density function}
\newacronym{pn}{PN}{phase noise}
\newacronym{pll}{PLL}{phase-locked loop}
\newacronym{pmf}{pmf}{probability mass function}
\newacronym{psd}{PSD}{power spectral density}
\newacronym{psk}{PSK}{phase-shift keying}
\newacronym{psr}{PSR}{pilot-to-signal power ratio}

\newacronym{qam}{QAM}{quadrature-amplitude modulation}

\newacronym{rhs}{r.h.s.}{right-hand side}
\newacronym{roadm}{ROADM}{reconfigurable optical add-drop multiplexer}
\newacronym{rp}{RP}{regular perturbation}
\newacronym{rrc}{RRC}{root-raised cosine}
\newacronym{rv}{RV}{random variable}

\newacronym{sdd}{SDD}{separate detection and decoding}
\newacronym{sdm}{SDM}{space-division multiplexing}
\newacronym{ssfm}{SSFM}{split-step Fourier method}
\newacronym{sic}{SD}{successive decoding}
\newacronym{simo}{SIMO}{single-in multiple-out}
\newacronym{sinr}{SINR}{signal to interference and noise ratio}
\newacronym{siso}{SISO}{single-in single-out}
\newacronym{sld}{SLD}{square-law detector}
\newacronym{snr}{SNR}{signal-to-noise ratio}
\newacronym{smf}{SMF}{single-mode fiber}
\newacronym{spa}{SPA}{sum-product algorithm}
\newacronym{spm}{SPM}{self-phase modulation}
\newacronym{ssb}{SSB}{single side-band}
\newacronym{ssbi}{SSBI}{signal-signal beat interference}
\newacronym{ssmf}{SSMF}{standard single-mode fiber}
\newacronym{svd}{SVD}{singular value decomposition}
\newacronym{svm}{SVM}{state vector machine}

\newacronym{te}{TE}{transverse electric}
\newacronym{tm}{TM}{transverse magnetic}
\newacronym{tp}{TP}{transparent propagation}

\newacronym{urr}{URR}{unidistant Rayleigh ring}

\newacronym{vamp}{VAMP}{vector approximate message passing}

\newacronym{wdm}{WDM}{wavelength-division multiplexing}
\newacronym{wlg}{w.l.g.}{without loss of generality}

\newacronym{xpm}{XPM}{cross-phase modulation}

\newacronym{zf}{ZF}{zero forcing}

\usepackage{amssymb} 
\usepackage{bm}

\usepackage{subfigure}
\usepackage{enumitem}
\usepackage[normalem]{ulem}

\usepackage{tikz}
\usetikzlibrary{positioning}
\usetikzlibrary{calc}
\usetikzlibrary{arrows}
\usepackage{pgfplots}
\pgfplotsset{compat=1.18}

\usepackage{algorithm}
\usepackage{algpseudocode}

\usepackage[detect-all]{siunitx}
\DeclareSIUnit\bpcu{bpcu}

\newcommand{\nsym}{{\ensuremath{n}}}

\newcommand{\ratiopilots}{{\ensuremath{\rho}}}

\renewcommand{\j}{{\ensuremath{\mathrm{j}}}}
\newcommand{\e}{{\ensuremath{\mathrm{e}}}}
\renewcommand{\vec}[1]{{\ensuremath{\bm{#1}}}}
\newcommand{\mat}[1]{{\ensuremath{\bm{#1}}}}
\newcommand{\transpose}{{\ensuremath{\mathsf{T}}}}
\newcommand{\hermitian}{{\ensuremath{\mathsf{H}}}}
\renewcommand{\v}[1]{\vec{#1}}
\newcommand{\m}[1]{\mat{#1}}
\newcommand{\real}[1]{{\mathrm{Re}\left\{#1\right\}}}

\newcommand{\secref}[1]{Sec.~\ref{#1}}
\newcommand{\figref}[1]{Fig.~\ref{#1}}

\tikzset{
    fnode/.style={
        draw, fill=black, minimum size = 10pt
        },
    vnode/.style={
        draw, shape = circle, minimum size = 25pt, inner sep = 0pt
        },
    msgarrow/.style={
        dashed, color = gray!60!black
        },
}

\definecolor{matlabBlue}{rgb}{0.00000,0.44700,0.74100}%
\definecolor{matlabRed}{rgb}{0.85000,0.32500,0.09800}%
\definecolor{matlabYellow}{rgb}{0.92900,0.69400,0.12500}%
\definecolor{matlabPurple}{rgb}{0.49400,0.18400,0.55600}%
\definecolor{matlabGreen}{rgb}{0.46600,0.67400,0.18800}%
\definecolor{matlabLightBlue}{rgb}{0.30100,0.74500,0.93300}%
\definecolor{matlabMagenta}{rgb}{0.63500,0.07800,0.18400}%

\definecolor{matlab1}{rgb}{0.00000,0.44700,0.74100}%
\definecolor{matlab2}{rgb}{0.85000,0.32500,0.09800}%
\definecolor{matlab3}{rgb}{0.92900,0.69400,0.12500}%
\definecolor{matlab4}{rgb}{0.49400,0.18400,0.55600}%
\definecolor{matlab5}{rgb}{0.46600,0.67400,0.18800}%
\definecolor{matlab6}{rgb}{0.30100,0.74500,0.93300}%
\definecolor{matlab7}{rgb}{0.63500,0.07800,0.18400}%

\definecolor{mycolor1}{rgb}{0.00000,0.44700,0.74100}%
\definecolor{mycolor2}{rgb}{0.85000,0.32500,0.09800}%
\definecolor{mycolor3}{rgb}{0.92900,0.69400,0.12500}%
\definecolor{mycolor4}{rgb}{0.49400,0.18400,0.55600}%
\definecolor{mycolor5}{rgb}{0.46600,0.67400,0.18800}%
\definecolor{mycolor6}{rgb}{0.30100,0.74500,0.93300}%
\definecolor{mycolor7}{rgb}{0.63500,0.07800,0.18400}%

\begin{document}
\title{Feedforward Phase Noise Compensation for Intersymbol Interference Channels}

\author{%
  \IEEEauthorblockN{Alex Jäger and Gerhard Kramer}
  \IEEEauthorblockA{
  School of Computation, Information and Technology,
  Technical University of Munich\\
    Email: \{alex.jaeger, gerhard.kramer\}@tum.de}

}

\maketitle

\begin{abstract}
A non-iterative phase noise compensation method based on the sum-product algorithm (SPA) is applied to the outputs of intersymbol interference (ISI) channels. The outputs are modeled as independent Gaussian random variables, and the receiver applies mismatched processing with von Mises statistics. The performance is compared with that of linear minimum-mean-square-error filtering. The SPA achieves higher information rates at similar complexity for three channel types: ISI-free, standard single-mode fiber, and multipath channels with orthogonal frequency-division multiplexing.
\end{abstract}

\newcommand\figscale{0.95}

\section{Introduction}
Oscillator imperfections cause \gls{pn} that distorts signals, e.g., generating \gls{ici} in \gls{ofdm} \cite{Sibel:18,Bowen:25}. We study \gls{pn} compensation for \gls{isi} channels, such as optical fibers with \gls{cd} and wireless links with multipath. We focus on discrete-time Wiener \gls{pn} processes that model free-running oscillators. The ideas extend to other \gls{pn} statistics with a Markov structure.

\begin{figure}[t!]
    \centering
    \scalebox{\figscale}{
    \begin{tikzpicture}
        \node (x) at (0,0) {$\vec{X}$};
        \node[draw, right = 1 of x, align=center] (isi) {\small ISI\\\small Channel};
        \node[draw, right = 1.25 of isi, align = center] (pn) {\small \gls{pn}\\\small Channel};
        \node[right = 1.5 of pn] (y) {$\vec{Y}$};
        \draw[->] (x)--(isi);
        \draw[->] (isi)--node[near start,above]{$\vec{Z}$}(pn);
        \draw[->] (pn)--(y);

        \node[circle, fill = black, right = 0.75 of pn, minimum size = 0.1, inner sep = 2pt] (dot) {};

        \node[draw, below = 1.5 of pn.center, align = center, anchor = center] (pncomp) {\small \gls{pn}\\\small Compen.};
        \node[draw, below = 1.5 of isi.center, align = center, anchor = center] (isicomp) {\small ISI\\\small Compen.};
        \node[below = 1.5 of x.center, anchor = center] (xhat) {$\v Y''$};

        \draw[->] (dot) |- (pncomp);
        \draw[->] (pncomp) -- node[near end,above] {$\;\v Y'$} (isicomp);
        \draw[->] (isicomp)--(xhat);

        \draw[->] ($(pncomp)+(0,-1)$) node[below]{$\v S$} -- (pncomp);

        \draw[dashed] ($(pn.north)+(-1.1,.2)$) -- node[midway,above]{$\v Y'\approx \v Z + \v W'$}($(pn.north)+(1.1,.2)$)--($(pncomp.south)+(1.1,-.2)$)--($(pncomp.south)+(-1.1,-.2)$)--($(pn.north)+(-1.1,.2)$);

        \draw[dashed] ($(isi.north)+(-1.1,0.85)$) -- node[midway,above]{$\v Y''\approx \v X + \v N'$}($(pn.north)+(1.2,0.85)$)--($(pncomp.south)+(1.2,-.3)$)--($(isicomp.south)+(-1.1,-.3)$)--($(isi.north)+(-1.1,0.85)$);
    \end{tikzpicture}
    }
    \caption{System model. The side information $\v S$ is based on either pilot symbols or symbols decoded in a previous stage of a successive decoding strategy.}
    \label{fig:systemmodel}
\end{figure}
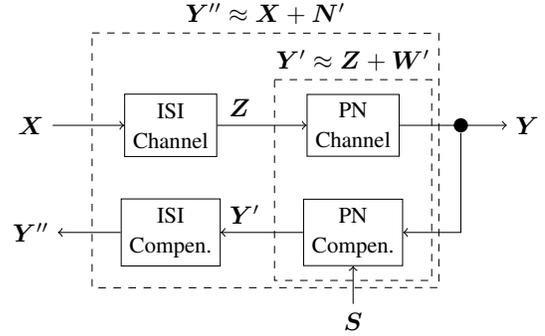

Consider the model in Fig.~\ref{fig:systemmodel} where $\v Z=[Z_1,\dots,Z_n]^\transpose$ is the output of a noisy \gls{isi} channel. The entries of $\v X$ may have discrete alphabets, but the entries of $\v Z$ are continuous due to the \gls{isi} taps and noise. Thus, \gls{pn} compensation algorithms such as Viterbi–Viterbi \cite{Viterbi:83} and blind phase search \cite{Pfau:09} do not apply since they require discrete-alphabet inputs. Depending on the design constraints, one may wish to perform \gls{pn} compensation after \gls{isi} compensation. However, this architecture enhances the \gls{pn} distortion and changes its statistics \cite{Shieh:08,Jung:25}.

We compensate \gls{pn} first, and consider two approaches. The first applies a \gls{lmmse} filter to $\v Y$ and outputs the phases \cite{Simon:GC01,Su:09}. The second calculates the \gls{map} decoding metrics $p_{Z_i|\v Y}$ with the \gls{spa} on a factor graph \cite{Kschischang:01}. This approach yields difficult integrals, and \cite{Dauwels:03,Dauwels:04} propose several approximations: quantize the phase, use surrogate Gaussian mixtures, numerical integration, particle filtering, or gradient methods. These algorithms exhibit tradeoffs between complexity and performance. Further simplifications represent the $p_{Z_i|\v Y}$ by their Fourier components or surrogates such as von Mises (Tikhonov) distributions to give closed-form integral expressions \cite{Colavolpe:05}. The papers \cite{Shayovitz:16,Kreimer:18,Szczecinski:20,Conti:24} refine the approximations and apply \gls{ep}. 

\subsection{Contributions and Organization}
We extend the \gls{spa}-based algorithms in \cite{Dauwels:03,Dauwels:04,Colavolpe:05,Shayovitz:16,Kreimer:18,Szczecinski:20,Conti:24} to \gls{isi} channels by modeling the $Z_1,\dots,Z_n$ as independent Gaussians. We study a feedforward structure without turbo iterations to avoid excessive latency. We also show that the \gls{spa} achieves higher information rates than the \gls{lmmse}-based algorithm of \cite{Simon:GC01} for a similar complexity.

This paper is organized as follows. \secref{sec:system_model} specifies the system model. \secref{sec:mismatched_processing} derives the decoding metrics, and \secref{sec:feedforward_compensation} shows how to compute them efficiently. \secref{sec:airs} shows how to bound \glspl{air} and reviews the \gls{lmmse} method of \cite{Simon:GC01}. \secref{sec:simulation-results} compares performance for three channels: \gls{isi}-free, \gls{ssmf}, and multipath. \secref{sec:conclusions} concludes the paper.

\emph{Notation.} Random variables and vectors are written in uppercase $X$ and $\v X$, and their realizations in lowercase $x$ and $\v x$. Uppercase bold $\v H$ also denotes a matrix. %
$\mathcal{N}(\v x;\v \mu,\m\Sigma)$ is the \gls{pdf} of a Gaussian with mean $\v\mu$ and covariance matrix $\m\Sigma$. $\mathcal{N}_{\mathbb{C}}(\v x;\v\mu,\m\Sigma)$ is the \gls{pdf} of a \gls{cscg} that is offset by $\v \mu$ (offset-\gls{cscg}). The von Mises (or Tikhonov) \gls{pdf} is
\begin{align}
    \mathcal{T}(\theta;\kappa) = \frac{1}{2\pi I_0(|\kappa|)}\exp\left(\real{\kappa\e^{-\j\theta}}\right)
\end{align}
where $I_0(\cdot)$ is the modified Bessel function of the first kind and zeroth order, and $\real{x}$ is the real part of $x$. The operator $\mathcal{P}\{\cdot\}$, which we specify in \eqref{eq:projection_channel}-\eqref{eq:projection} and \eqref{equ:recursion_alpha} below, projects onto a family of distributions.

\section{System Model}
\label{sec:system_model}

Consider $\v X=[X_1,\ldots,X_\nsym]^\transpose$ with \gls{iid} entries. The average transmit power is $\nu_x=\mathrm{E}[||\v X||_2^2]/\nsym$. Consider a complex \gls{isi} channel with \gls{awgn}:
\begin{align}
    \v Z =\m H\v X+\v N
\end{align}
where $\v N\sim\mathcal{N}_{\mathbb{C}}(\v 0,\nu_n \m I )$. For example, optical fiber has \gls{cd} with unitary $\m H$, and \gls{ofdm} systems effectively have
\begin{align}
\label{eq:ofdm_channel}
    \m H = \m H_c\m F^\hermitian
    = \m F^\hermitian \m \Delta    
\end{align}
where $\m H_c=\m F^\hermitian\m\Delta\m F$ is circulant, $\m F$ is a \gls{dft} matrix, and $\m\Delta$ is diagonal. Note that the \gls{ofdm} model neglects the cyclic prefix that can improve the \gls{pn} compensation; see Sec.~\ref{subsec:multi-path-channel}. Note also that the $Z_i$ are approximately \gls{cscg} for many \gls{isi} channel taps, or more generally if $\m H$ is dense as for OFDM or other precoders. This is different from \gls{isi}-free models with discrete channel inputs \cite{Dauwels:04,Colavolpe:05,Conti:24,Shayovitz:16,Kreimer:18,Szczecinski:20}. 

The vector $\v Z$ is fed to a \gls{pn} channel
\begin{align}
    \label{eq:pn_channel}
    \v Y = \v Z \odot \e^{\j \v\Theta} + \v W,
    \quad \v W\sim\mathcal{N}_{\mathbb{C}}(\v0,\nu_w\m I)
\end{align}
where $e^{\j \v x}$ is a vector with entries $e^{\j x_i}$ and $\odot$ denotes a term-by-term product. Discrete-time Wiener \gls{pn} has
\begin{align}
    \Theta_i = \Theta_{i-1}+\Delta_i \mod 2\pi
    \label{eq:dt-Wiener-PN}
\end{align}
where the $\{\Delta_i\}_{i=1}^\nsym$ are independent, zero-mean, Gaussian increments with variance $\nu_\Delta$. The initial phase $\Theta_0$ is uniformly distributed in $[-\pi,\pi)$ and unknown to the transceivers. 

We design a \gls{pn} compensator that outputs an equalized
\begin{align}
    \v Y' \approx \v Z+\v W'
    \label{eq:pn_compensation_surrogate}
\end{align}
with $\v W'\sim\mathcal{N}_{\mathbb{C}}(\v 0, \nu_{w'}\m I )$ and small $\nu_{w'}$; see \figref{fig:systemmodel}.
The $\v Y'$ are processed by an \gls{isi} compensation module to give
\begin{align}
    \v Y'' \approx \v X+\v N'
    \label{eq:isi_compensation_surrogate}
\end{align}
with $\v N'\sim\mathcal{N}_{\mathbb{C}}(\v 0, \m \Sigma_{n'})$. For example, for unitary $\m H$ and \gls{ofdm} we apply the respective linear equalizers
\begin{align}
    \v Y'' = \m H^\hermitian\v Y', \quad
    \v Y'' = \m F\v Y' .
    \label{eq:linear-equalizer}
\end{align}
For both cases, we have $\m \Sigma_{n'} = (\nu_n+\nu_{w'})\m I$.

\subsection{Side Information}
\label{subsec:side_information}
Let $\v X = \v P + \v M$, where $\v P$ and $\v M$ are vectors of pilot and message symbols, respectively. We measure the pilot overhead via the \gls{psr} $\ratiopilots=\nu_p/\nu_x$ where $\nu_p=\mathrm{E}[||\v P||_2^2]/\nsym$. 
Let $\v S=\m H \v P$ and $\v T=\m H \v M$ so that
\begin{align}
    \label{eq:pilots_isi}
    \v Z = \v S + \v T + \v N.
\end{align}

Note that the side information in Fig.~\ref{fig:systemmodel} can model message symbols decoded in previous decoding stages when using \gls{sic} \cite{Guess:00:ISI,Pfister:01,prinz2023successive,Jaeger-JSAC25}.

\section{Mismatched Processing}
\label{sec:mismatched_processing}
The \gls{map} estimate of $\v X$ with side information $\v s$ is
\begin{align}
    \hat{\v x} = \arg\max_{\v x} P(\v x|\v s)\, p(\v y|\v x) 
\end{align}
where $P(\v x|\v s)$ carries \emph{a priori} information and
\begin{align}
    & p(\v y|\v x)
    = \int\nolimits_{\mathbb{C}^\nsym}
    p(\v z|\v x)\, p(\v y|\v z)\, \mathrm{d}\v z
    \nonumber \\
    & = \int\nolimits_{\mathbb{C}^\nsym} p(\v z|\v x) 
    \left(
    \int\nolimits_{[-\pi,\pi)^\nsym} p(\v \theta)\, p(\v y|\v z,\v \theta)\, \mathrm{d}\v\theta
    \right) \mathrm{d}\v z.
    \label{eq:channel_probability}
\end{align}
Computing \eqref{eq:channel_probability} for all $\v x$ has exponential complexity in $\nsym$.

We reduce complexity by modeling the $T_i= (\m H \v M)_i$ as offset-\gls{cscg} and independent via the surrogate model
\begin{align}
    q(\v z |\v s)
    = \mathcal{N}_{\mathbb{C}}(\v z;\v s, (\nu_t+\nu_n)\m I)
    \label{eq:surrogat_prior_t}
\end{align}
where $\nu_t=\mathrm{E}[||\v T||_2^2]/\nsym$. 
We further simplify by computing
\begin{align}
    q(\v y|\v z, \v s)
    = \frac{\mathcal{P}\left\{q(\v z|\v s)p(\v y|\v z)\right\}}{q(\v z|\v s)}
    \label{eq:projection_channel}
\end{align}
where $\mathcal{P}\{\cdot\}$ projects onto a multivariate offset-\gls{cscg} with a scaled identity covariance matrix:
\begin{align}
    \mathcal{P}\left\{q(\v z|\v s)p(\v y|\v z)\right\} = \mathcal{N}_{\mathbb{C}}(\v z;\hat{\v z},\nu\m I).
    \label{eq:projection}
\end{align}
We use the \gls{ep} approach and minimize the \gls{id}, which simplifies to moment matching:
\begin{align}
    \hat{\v z} &= \mathrm{E}[\v Z | \v Y=\v y,\v S = \v s] \label{eq:zi} \\
    \nu &= \frac{1}{\nsym}\sum\nolimits_{i=1}^\nsym\mathrm{Var}[Z_i | \v Y=\v y, \v S = \v s]
    \label{eq:nu}
\end{align}
where the expectations are based on the surrogate posterior
\begin{align}
    \frac{q(\v z|\v s)p(\v y|\v z) }{ \int_{\mathbb C^n} q(\v z'|\v s)\, p(\v y|\v z')\, \mathrm{d}\v z' } .
    \label{eq:normalization}
\end{align}
\secref{sec:feedforward_compensation} uses message passing to calculate the moments. Note that $\hat{\v z}$ is a \gls{mmse} estimate. 

The surrogate channel \eqref{eq:projection_channel} is thus (see \cite{Bromiley:14})
\begin{align}
    q(\v y' | \v z,\v s)
    = \frac{\mathcal{N}_{\mathbb{C}}(\v z;\hat{\v z},\nu\m I)}{\mathcal{N}_{\mathbb{C}}(\v z;\v s, (\nu_t+\nu_n)\m I)}
    \propto \mathcal{N}_{\mathbb{C}}(\v y'; \v z,\nu_{w'}\m I)
    \label{eq:yprime_surrogate_channel}
\end{align}
where we switched notation from $\v y$ to $\v y'$ to emphasize the relation to $\v Y'$ in \eqref{eq:pn_compensation_surrogate}, and where
\begin{align}
    \v y' &= \frac{\hat{\v z}(\nu_t+\nu_n)-\v s\nu}{\nu_t+\nu_n-\nu},\quad    \nu_{w'} = \frac{\nu(\nu_t+\nu_n)}{\nu_t+\nu_n-\nu}. \label{eq:extrinsic}
\end{align}
The channel from $\v z$ to $\v y'$ is thus described by an \gls{awgn} channel as in \eqref{eq:pn_compensation_surrogate}. The \gls{pn} compensation outputs a real scalar (a variance) and a complex vector of dimension $\nsym$ (the means). 

We remark that using a linear filter for \gls{isi} compensation as in \eqref{eq:linear-equalizer} is the same as calculating (see \eqref{eq:channel_probability})
\begin{equation}
    \begin{aligned}
        q(\v y''|\v x,\v s)
        & = \int_{\mathbb{C}^\nsym} p(\v z| \v x)\, q(\v y'' | \v z,\v s)\, \mathrm{d}\v z \\
        & = \mathcal{N}_{\mathbb{C}}(\v y'';\m \Sigma \v x,(\nu_n+\nu_{w'})\m I)
    \end{aligned}
    \label{eq:yprimeprime_metric}
\end{equation}
where we switched notation from $\v y'$ to $\v y''$ to emphasize the relation to $\v Y''$ in  \eqref{eq:linear-equalizer}. We have $\m\Sigma=\m I$ for \gls{cd} and $\m \Sigma=\m\Delta$ for \gls{ofdm}.

\section{SPA \Gls{pn} Compensation}
\begin{figure}[t!]
    \centering
    \scalebox{\figscale}{
    \begin{tikzpicture}

    \node[fnode, label=above: \small$q(z_{i-1}|s_{i-1})$] (pm) at (0,0) {};
    \node[fnode, label=above: \small$q(z_{i}|s_{i})$, right = 2 of pm.center, anchor = center] (p) {};
    \node[fnode, label=above: \small$q(z_{i+1}|s_{i+1})$, right = 2 of p.center, anchor = center] (pp) {};

    \node[vnode, below = .5 of pm] (Xm) {\small$z_{i-1}$};
    \node[vnode, below = .5 of p] (X) {\small$z_{i}$};
    \node[vnode, below = .5 of pp] (Xp) {\small$z_{i+1}$};

    \node[fnode, label = left: $f_{i-1}$, below = .5 of Xm] (fm) {};
    \node[fnode, label = left: $f_i$, below = .5 of X] (f) {};
    \node[fnode, label = left: $f_{i+1}$, below = .5 of Xp] (fp) {};

    \node[vnode, below = 1 of fm] (Thetam) {\small$\theta_{i-1}$};
    \node[vnode, below = 1 of f] (Theta) {\small$\theta_{i}$};
    \node[vnode, below = 1 of fp] (Thetap) {\small$\theta_{i+1}$};

    \node[fnode, label = below: \small$p(\theta_i|\theta_{i-1})$, below right = 1 and 1 of Thetam.south, anchor = north] (qm) {};
    \node[fnode, label = below: \small$p(\theta_{i+1}|\theta_{i})$, below right = 1 and 1 of Theta.south, anchor = north] (q) {};

    \draw (pm)--(Xm)--(fm)--(Thetam);
    \draw (p)--(X)--(f)--(Theta);
    \draw (pp)--(Xp)--(fp)--(Thetap);

    \draw (Thetam)--(qm)--(Theta)--(q)--(Thetap);
    \draw[dashed] (Thetam)--++(-.75,-1.5);
    \draw[dashed] (Thetap)--++(.75,-1.5);

    \draw[->, dashed] (f) edge[bend right] node[midway,left] {$\gamma_i$} (Theta);
    \draw[->, dashed] (qm) edge[bend left] node[midway,left] {$\alpha_i$} (Theta);
    \draw[->, dashed] (q) edge[bend right] node[midway,right] {$\beta_i$} (Theta);
        
\end{tikzpicture}
    }
    \caption{Factor graph of the \gls{pn} compensation. Circle nodes represent variables and square nodes represent functions. The node $f_i$ represents $p(y_i|z_i,\theta_i)$.}
    \label{fig:factorGraph}
\end{figure}
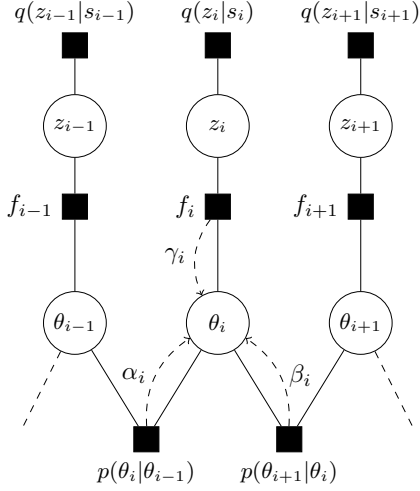

Fig.~\ref{fig:factorGraph} shows the factor graph of the \gls{pn} compensation. The graph has a tree structure and requires only one forward and backward pass. We focus on discrete-time Wiener \gls{pn}, but the methods generalize to other \gls{pn} statistics with a Markov structure by using Gaussian or von Mises surrogate priors.
\label{sec:feedforward_compensation}

Consider the surrogate prior \eqref{eq:surrogat_prior_t} and define $\nu_{z|s}=\nu_t+\nu_n$. To calculate $\hat{\v z}$ and $\nu$ in \eqref{eq:zi} and \eqref{eq:nu},
the \gls{spa} calculates
\begin{align}
    \gamma_i(\theta_i) &= \int_\mathbb{C} q(z_i|s_i)\, p(y_i|z_i,\theta_i)\, \mathrm{d} z_i
    \nonumber \\
    & = \mathcal{N}_\mathbb{C} \left( y_i ; s_i\e^{\j\theta_i}, \nu_{z|s}+\nu_w \right)
    \propto \mathcal{T} \left(\theta_i; \kappa_{\gamma,i} \right)
    \label{eq:gammma}
\end{align}
with
$\kappa_{\gamma,i}=2y_i s_i^*/(\nu_{z|s}+\nu_w)$, and $\gamma_i$ is proportional to a von Mises density in $\theta_i$. This differs from discrete inputs where $\gamma_i(\theta_i)$ is a Gaussian mixture. This mixture can be projected onto a von Mises density \cite[Sec.~IV-B]{Colavolpe:05} and the projection can be improved using techniques such as \gls{ep} \cite{Shayovitz:16,Kreimer:18,Szczecinski:20,Conti:24}. As a result, the algorithms in \cite{Colavolpe:05,Shayovitz:16,Kreimer:18,Szczecinski:20,Conti:24} would match ours but with a different starting point.

The forward and backward paths compute the recursions
\begin{align}
    \alpha_{i}(\theta_{i}) &\propto \int_{-\pi}^\pi \alpha_{i-1}(\theta_{i-1})\, \gamma_{i-1}(\theta_{i-1})\, p(\theta_{i}|\theta_{i-1})\, \mathrm{d}\theta_{i-1}
    \label{eq:forwardRecursion} \\
    \beta_{i}(\theta_{i}) &\propto \int_{-\pi}^\pi\beta_{i+1}(\theta_{i+1})\, \gamma_{i+1}(\theta_{i+1})\, p(\theta_{i+1}|\theta_{i})\, \mathrm{d}\theta_{i+1}
    \label{eq:backwardRecursion}    
\end{align}
with $\alpha_1(\theta_1)  \propto 1$, and $\beta_\nsym(\theta_\nsym)\propto 1$. The posterior reads
\begin{align}
    q(z_i|\vec{y},\v s)
    \propto q(z_i|s_i)\int_{-\pi}^\pi \alpha_i(\theta_i)\, \beta_i(\theta_i)\, p(y_i|z_i,\theta_i)\, \mathrm{d}\theta_i.
        \label{eq:posterior}
\end{align}
These integrals are difficult to compute, so we project onto von Mises densities. Recall from \eqref{eq:gammma} that $\gamma_i$ is considered von Mises in $\theta_i$, and \eqref{eq:posterior} is easy to solve if $\alpha_i$ and $\beta_i$ are von Mises in $\theta_i$. We thus project $\alpha_i$ and $\beta_i$ to von Mises densities after each step of the recursions \eqref{eq:forwardRecursion} and \eqref{eq:backwardRecursion}. The \glspl{id} are minimized by matching circular means \cite[App. A]{Szczecinski:20}. 

The recursions are now (focusing on time $i+1$)
\begin{align}  
    \alpha_{i+1}(\theta_{i+1}) &=\mathcal{P}\left\{ \int_{-\pi}^\pi \alpha_i(\theta_i)\, \gamma_i(\theta_i)\, p(\theta_{i+1}|\theta_i)\, \mathrm{d}\theta_i\right\} \nonumber \\
    & = \mathcal{T}(\theta_{i+1},\kappa_{\alpha,i+1}) \label{equ:recursion_alpha}
\end{align}
and likewise for $\beta_{i-1}(\theta_{i-1})$. The product $\alpha_i(\theta_i)\,\gamma_i(\theta_i)$ is von Mises. Convolving with the wrapped Gaussian $p(\theta_{i+1}|\theta_i)$ gives a density whose circular mean is the product of the circular means of $\alpha_i(\theta_i)\gamma_i(\theta_i)$ and $p(\theta_{i+1}|\theta_i)$ \cite[Lemma 3]{Kurz:16}. Calculating $\kappa_{\alpha,i+1}$ from the circular mean requires inverting a ratio of Bessel functions. The paper \cite[App. C]{Szczecinski:20} approximates the inverse for small $\nu_\Delta$ and $|\kappa_{\alpha,i}+\kappa_{\gamma,i}|>1$. Both conditions are satisfied in our setups. We thus use the recursion proposed in \cite[Sec. IV-A]{Szczecinski:20} which coincides with \cite[Sec. IV-B]{Colavolpe:05}: set $\kappa_{\alpha,1}=\kappa_{\beta,{\nsym}}=0$ and compute
\begin{subequations}
\begin{align}
    \kappa_{\alpha,{i+1}} &= \frac{\kappa_{\alpha,i}+\kappa_{\gamma,i}}{1+\nu_\Delta |\kappa_{\alpha,i}+\kappa_{\gamma,i}|}, \quad i=1,\ldots,\nsym-1
    \label{eq:forward_path} \\
    \kappa_{\beta,{i-1}} &= \frac{\kappa_{\beta,i}+\kappa_{\gamma,i}}{1+\nu_\Delta |\kappa_{\beta,i}+\kappa_{\gamma,i}|}, \quad i=\nsym,\ldots,2 .
    \label{eq:backward_path}
\end{align}
\end{subequations}

With von Mises $\alpha_i$ and $\beta_i$, the posterior \eqref{eq:posterior} is
\begin{align}    
    q(z_i|\v y,\v s)
    & = q(z_i|s_i) \frac{\nu_{z|s}+\nu_w}{\nu_w}\cdot\nonumber\\  &
    \frac{\exp\left(-\frac{|y_i|^2+|z_i|^2}{\nu_w}\right)}{\exp\left(-\frac{|y_i|^2+|s_i|^2}{\nu_{z|s}+\nu_w}\right)} \frac{I_0\left(\left|\kappa_{\alpha,i}+\kappa_{\beta,i}+2\frac{y_iz_i^*}{\nu_w}\right|\right)}{I_0\left(\left|\kappa_{\alpha,i}+\kappa_{\beta,i}+\kappa_{\gamma,i}\right|\right)}.
    \label{eq:posterior_cf}
\end{align}
Finally, we project onto a Gaussian to obtain
\begin{subequations}
\label{eq:cscg_posterior_mv}
\begin{align}
    \hat z_i &= s_i\frac{\nu_w}{\nu_w+\nu_{z|s}}+y_i\frac{\nu_{z|s}}{\nu_w+\nu_{z|s}}\frac{I_1(|\kappa_i|)}{I_0(|\kappa_i|)}\e^{-\j\angle\kappa_i}
    \label{eq:cscg_posterior_mean}
    \\
    \nu &= \frac{\nu_w\nu_{z|s}}{\nu_w+\nu_{z|s}}+\frac{\nu_{z|s}^2}{\nsym(\nu_w+\nu_{z|s})^2}
    \sum_{i=1}^\nsym|y_i|^2\left(1-\frac{I_1(|\kappa_i|)^2}{I_0(|\kappa_i|)^2}\right)
    \label{eq:cscg_posterior_variance}
\end{align}
\end{subequations}
where $\kappa_i=\kappa_{\alpha,i}+\kappa_{\beta,i}+\kappa_{\gamma,i}$. Finally, compute $\v y'$ via \eqref{eq:extrinsic}.

The $\kappa_{\gamma,i}$, $\hat z_i$, and $\nu$ can be computed in parallel. The metrics in \eqref{eq:forward_path} and \eqref{eq:backward_path} are computed sequentially. In total, the receiver requires $27\nsym+3$ multiplications, $16\nsym+2$ additions, and $5\nsym$ \gls{lut} accesses for nonlinear functions.

\section{Achievable Information Rates}
\label{sec:airs}
The \gls{gmi} \cite{Scarlett:20,kramer:23}
\begin{align}
    \label{eq:rate}
    I_q(\v X; \v Y'' | \v S) = 
    \mathrm{E}\left[\log\frac{q(\v Y''|\v X,\v S)}{q(\v Y''|\v S)} \right]
\end{align}
gives an \gls{air}, where $q(\v y''|\v x,\v s)$ is a surrogate channel and
\begin{align}
    q(\v y''|\v s) =
    \int_{\mathbb{C}^\nsym} p(\v x|\v s)\,q(\v y''|\v x,\v s)\, \mathrm{d}\v x.
\end{align}
A \gls{jdd} surrogate $q(\v y''|\v x,\v s)$ includes dependencies among the entries of $\v x$ and $\v y''$. A \gls{sdd} surrogate ignores these; we use the Gaussian surrogate \eqref{eq:yprimeprime_metric}.

\subsection{Rate Upper Bounds}
Consider the \gls{isi}-free channel (see~\eqref{eq:pn_channel} and~\eqref{eq:pilots_isi})
\begin{align}
    \v Y = (\v S + \v T + \v N) \odot \e^{\j\v\Theta} + \v W
    \label{eq:ISI-free-channel}
\end{align}
with \gls{iid} \gls{cscg} inputs $\v S$ and $\v T$. A \gls{jdd} rate upper bound can be based on the \gls{mi}
\begin{align}
    \label{eq:jdd_receiver_air}
    I(\v X;\v Y|\v S) \le I(\v Z;\v Y|\v S)
    = I(\v Z;\v Y) - I(\v S;\v Y)
\end{align}
where we used the Markov chain $\v S-\v X-\v Z-\v Y$. The inequality in \eqref{eq:jdd_receiver_air} is an equality if $\nu_n=0$ and the channel $\m H$ is invertible for the support of $\v X$.
We have
\begin{align}
    (\v T + \v N)\e^{\j\v \Theta}+\v W
    \sim \mathcal{N}_{\mathbb{C}}(\v 0,(\nu_t+\nu_n+\nu_w)\m I)
\end{align}
and the terms $I(\v Z;\v Y)$ and $I(\v S;\v Y)$ correspond to Wiener \gls{pn} channels with the respective \glspl{snr}
$(\nu_{s}+\nu_{t}+\nu_{n})/\nu_w$ and 
$\nu_{s}/(\nu_{t}+\nu_{n}+\nu_w)$.

We write $\overline{I}(\v Z;\v Y|\v S)$ and $\underline{I}(\v Z;\v Y|\v S)$ as upper \cite{Barletta:12} and lower \cite{Dauwels:08} bounds on \eqref{eq:jdd_receiver_air}, respectively; see also \cite{Lapidoth:02,Khanzadi:15}.
The \gls{sdd} \gls{air} is based on the \gls{sdd} metrics $p(z_i|\v y,\v s)$:
\begin{align}
\label{eq:sdd_receiver_air}
    \frac{1}{\nsym}\sum_{i=1}^\nsym I(Z_i;\v Y|\v S)
    = h(T+N) - \frac{1}{\nsym}\sum_{i=1}^\nsym h(Z_i|\v Y,\v S)
\end{align}
with $h(T+N)=\log(\pi\e(\nu_{t}+\nu_n))$. We quantize the \gls{pn} and run the \gls{spa} to compute upper bounds on $h(Z_i|\v Y,\v S)$ that give lower bounds on \eqref{eq:sdd_receiver_air}; see \cite{Barletta:12}. We increased the quantization precision until we saw no more improvement in rate. The lower bound is thus a good estimate of the actual \gls{mi}.

\subsection{\gls{lmmse} Rates and Complexity}
The \gls{lmmse} filters in \cite{Simon:GC01} are computed via the matrix form
\begin{align}
    \v V = \mathrm{E}\left[ \left(\v s \odot \e^{\j(\v \Theta-\angle \v s)}\right) \tilde{\v Y}^\hermitian \right]
    \mathrm{E}\left[ \tilde{\v Y} \tilde{\v Y}^\hermitian \right]^{-1}
    \label{eq:LMMSE-est}
\end{align}
where $\tilde{\v Y} = \v Y \odot \e^{-\j\angle \v s}$, and $\angle\v x$ has entries the phases of $x_i$. The expectation uses the mismatched prior \eqref{eq:surrogat_prior_t} and the phase-corrected output \eqref{eq:pn_compensation_surrogate} is $\v Y'=\v Y \odot \exp(-\j \angle (\v V\tilde{\v Y}))$.

The \gls{spa} has a complexity of $48$ operations per information symbol, where we count multiplications, additions, and \gls{lut} accesses equally. The \gls{lmmse} filter operates on complex numbers, requiring $4$ operations per filter tap. We limit the filter length to $25$ taps, which is approximately twice as complex as the \gls{spa}. Observe that the frequency-domain overlap-add method for convolution has similar complexity as direct time-domain convolution for such filter lengths.\footnote{Overlap-add would use 128-point \glspl{fft} here. The Cooley-Tukey \gls{fft} then requires $\approx 80$ operations per information symbol for the convolution, which is similar to a time-domain convolution.}

\section{Simulation Results}
\label{sec:simulation-results}

We study three channel types: \gls{isi}-free, \gls{ssmf} with \gls{cd}, and multipath with \gls{ofdm}. We consider two \gls{pn} scenarios with $(\text{SNR},\nu_\Delta)$ being $(\SI{5}{\dB},10^{-6})$ and $(\SI{13}{\dB},5\cdot 10^{-3})$.
We use two types of time-domain pilots with a fixed positive value: interleaved and superposed pilots. We found that superposed pilots achieve higher rates than interleaved pilots, and thus focus mainly on them. We simulate with $\nu_n=0$ and by averaging over $2^8$ sequences of length $\nsym=2^{13}$ for all curves.

\subsection{ISI-free Channel}
\label{subsec:isi-free}
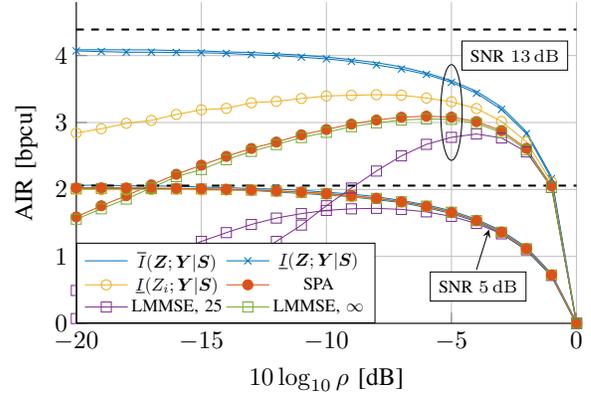
\begin{figure}[t!]
    \centering
    \scalebox{\figscale}{
    \begin{tikzpicture}
    \begin{axis}[%
    width=7cm,
    height=4.5cm,
    at={(0,0)},
    scale only axis,
    xmin=-20,
    xmax=0,
    xlabel={$10\log_{10}\ratiopilots$ [dB]},
    ymin=0,
    ymax=4.8,
    ylabel={AIR [bpcu]},
    axis x line*=bottom,
    axis y line*=left,
    ytick={0,1,...,5},
    xmajorgrids,
    ymajorgrids,
    legend style={at={(0,0)},anchor=south west,nodes={scale=0.7, transform shape}},
    legend columns = 2,
    ]        
        \addplot [color=mycolor1]
          table[]{figures/data/13dB_5e-3_air_v_k_jddupper.tsv};
            \addlegendentry{$\overline{I}(\v Z;\v Y|\v S)$};
        
        \addplot [color=mycolor1, mark = x]
          table[]{figures/data/13dB_5e-3_air_v_k_jddlower.tsv};
            \addlegendentry{$\underline{I}(\v Z;\v Y|\v S)$};
        \addplot [color=mycolor3, mark = o]
          table[]{figures/data/13dB_5e-3_air_v_k_sdd.tsv};
            \addlegendentry{$\underline{I}(Z_i;\v Y|\v S)$};
        \addplot [color=mycolor2, mark = *]
          table[]{figures/data/13dB_5e-3_air_v_k_cscg.tsv};
            \addlegendentry{SPA};
        \addplot [color=mycolor4, mark = square]
          table[]{figures/data/5dB_1e-6_air_v_k_LMMSE_L_50.tsv};
            \addlegendentry{LMMSE, 25};
        \addplot [color=mycolor5, mark = square]
          table[]{figures/data/5dB_1e-6_air_v_k_LMMSE.tsv};
            \addlegendentry{LMMSE, $\infty$};

        \draw[thick,dashed] (axis cs: -20,4.3891)--(axis cs: 0,4.3891);

        \addplot [color=mycolor1, forget plot]
          table[]{figures/data/5dB_1e-6_air_v_k_jddupper.tsv};
        \addplot [color=mycolor1, mark = x, forget plot]
          table[]{figures/data/5dB_1e-6_air_v_k_jddlower.tsv};
        \addplot [color=mycolor3, mark = o, forget plot]
          table[]{figures/data/5dB_1e-6_air_v_k_sdd.tsv};
        \addplot [color=mycolor2, mark = *, forget plot]
          table[]{figures/data/5dB_air_v_k_cscg.tsv};
        \addplot [color=mycolor4, mark = square, forget plot]
          table[]{figures/data/13dB_5e-3_air_v_k_LMMSE_L_12.tsv};
        \addplot [color=mycolor5, mark = square, forget plot]
          table[]{figures/data/13dB_5e-3_air_v_k_LMMSE.tsv};

        \draw[thick,dashed] (axis cs: -20,2.0574)--(axis cs: 0,2.0574);

        \node[draw,right,fill=white,align=left] at (axis cs: -4.6,4) {\scriptsize SNR $\SI{13}{\dB}$};
        \draw (axis cs: -5,3.15) ellipse (0.15cm and 0.67cm);
        \node[draw,left,fill=white,align=left] at (axis cs: -2,0.55) {\scriptsize SNR $\SI{5}{\dB}$};
        \draw[-stealth] (axis cs: -4,0.8) -- (axis cs: -3.5,1.4);
        
    \end{axis}
\end{tikzpicture}%
    }
    \caption{Rates for \gls{isi}-free channels with $(\text{SNR},\nu_\Delta)=(\SI{13}{\dB},5\cdot 10^{-3})$ (upper curves) and $(\text{SNR},\nu_\Delta)=(\SI{5}{\dB},10^{-6})$ (lower curves).}
    \label{fig:air_snr}
\end{figure}
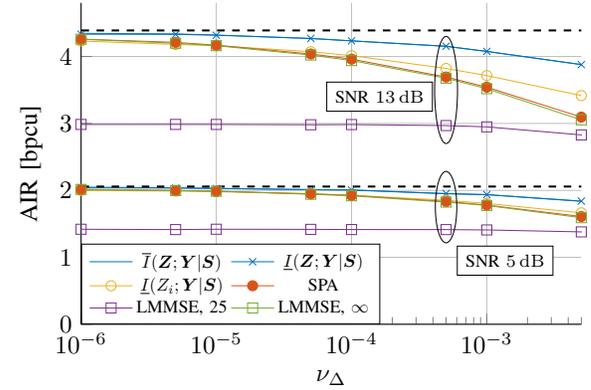
\begin{figure}[t!]
    \centering
    \scalebox{\figscale}{
    \begin{tikzpicture}
    \begin{axis}[%
    width=7cm,
    height=4.5cm,
    at={(0,0)},
    scale only axis,
    xmode=log,
    xmin = 0.000001,
    xmax = 0.005,
    xlabel={$\nu_\Delta$},
    ymin=0,
    ymax=4.8,
    ylabel={AIR [bpcu]},
    axis x line*=bottom,
    axis y line*=left,
    ytick={0,1,...,5},
    xmajorgrids,
    ymajorgrids,
    legend style={at={(0,0)},anchor=south west,nodes={scale=0.7, transform shape}},
    legend columns = 2,
    ];
        
        \addplot [color=mycolor1, mark = s]
          table[]{figures/data/5dB_air_v_nuDelta_jddupper.tsv};
        \addlegendentry{$\overline{I}(\v Z;\v Y|\v S)$};
        
        \addplot [color=mycolor1, mark = x]
          table[]{figures/data/5dB_air_v_nuDelta_jddlower.tsv};
        \addlegendentry{$\underline{I}(\v Z;\v Y|\v S)$};
        
        \addplot [color=mycolor3, mark = o]
          table[]{figures/data/5dB_air_v_nuDelta_sdd.tsv};
        \addlegendentry{$\underline{I}(Z_i;\v Y|\v S)$};
        
        \addplot [color=mycolor2, mark = *]
          table[]{figures/data/5dB_air_v_nuDelta_cscg.tsv};
        \addlegendentry{SPA};

        \addplot [color=mycolor4, mark = square]
          table[]{figures/data/5dB_air_v_nuDelta_LMMSE_L_12.tsv};
        \addlegendentry{LMMSE, 25};

        \addplot [color=mycolor5, mark = square]
          table[]{figures/data/5dB_air_v_nuDelta_LMMSE.tsv};
        \addlegendentry{LMMSE, $\infty$};

        \addplot [color=mycolor1, mark = s, forget plot]
          table[]{figures/data/13dB_air_v_nuDelta_jddupper.tsv};
        
        \addplot [color=mycolor1, mark = x, forget plot]
          table[]{figures/data/13dB_air_v_nuDelta_jddlower.tsv};
        
        \addplot [color=mycolor3, mark = o, forget plot]
          table[]{figures/data/13dB_air_v_nuDelta_sdd.tsv};
        
        \addplot [color=mycolor2, mark = *, forget plot]
          table[]{figures/data/13dB_air_v_nuDelta_cscg.tsv};

        \addplot [color=mycolor4, mark = square, forget plot]
          table[]{figures/data/13dB_air_v_nuDelta_LMMSE_L_12.tsv};

      \addplot [color=mycolor5, mark = square, forget plot]
          table[]{figures/data/13dB_air_v_nuDelta_LMMSE.tsv};
            
        \draw[thick,dashed] (axis cs: 0.0000001,4.3891)--(axis cs: 0.005,4.3891);

        \draw[thick,dashed] (axis cs: 0.0000001,2.0574)--(axis cs: 0.005,2.0574);

        \node[draw,right,fill=white,align=right] at (axis cs: 0.000065,3.4) {\scriptsize SNR $\SI{13}{\dB}$};
        \draw (axis cs: 0.0005,3.5) ellipse (0.16cm and 0.75cm);
        \node[draw,right,fill=white,align=left] at (axis cs: 0.0006,0.95) {\scriptsize SNR $\SI{5}{\dB}$};
        \draw (axis cs: 0.0005,1.75) ellipse (0.15cm and 0.5cm);
    \end{axis}
\end{tikzpicture}%
    }
    \caption{Rates for \gls{isi}-free channels and optimized $\rho$ with $\text{SNR}=\SI{13}{\dB}$ (upper curves) and $\text{SNR}=\SI{5}{\dB}$ (lower curves).} 
    \label{fig:air_nu_delta}
\end{figure}

Consider the \gls{isi}-free channel \eqref{eq:ISI-free-channel} and superposed pilots. \figref{fig:air_snr} plots the \glspl{air} of \gls{cscg} modulation against the \gls{psr} $\ratiopilots=\nu_p/\nu_x$ in dB. The horizontal dashed lines show the \glspl{air} of \gls{awgn} channels with the same \glspl{snr}. The \gls{spa} achieves the best \gls{sdd} rates at low \gls{snr} (lower curves), and the optimal $\rho$ is small. However, there are gaps at high \gls{snr} and \gls{pn} variance (upper curves). Tracking the \gls{pn} improves with $\rho$, but the transmit entropy decreases. The best tradeoff is at $\ratiopilots\!\approx\!\frac{1}{3}$ or \SI{-5}{\dB}, where the SPA rate is $\approx\:$\SI{80}{\percent} of the \gls{air} $I(\v Z;\v Y|\v S)$. The purple and green lines show the \gls{lmmse} \glspl{air}; the full \gls{lmmse} rates and SPA rates almost coincide.

\figref{fig:air_nu_delta} plots the \glspl{air} against the \gls{pn} variance $\nu_\Delta$, where we optimized $\ratiopilots$ to maximize the SPA rate. This $\ratiopilots$ was used to compute the \gls{sdd} rate $\underline{I}(Z_i;\v Y|\v S)$ and \gls{jdd} rate $\underline{I}(\v Z;\v Y|\v S)$.

\subsection{Fiber-Optic Channel}
\label{subsec:fiber-optic-channel}
Figs. \ref{fig:AIR_v_k_CD_uniform} and \ref{fig:AIR_v_k_CD_shaped}
plot the \glspl{air} against the \gls{psr} for \SI{10}{\kilo\meter} of \gls{ssmf} without nonlinearity. The first equalizer in \eqref{eq:linear-equalizer} is applied to $\v Y'$. The message symbols $M_i$ have a \gls{qam} alphabet and a uniform distribution. We see that the 64-QAM rates with superposed pilots approach the \glspl{air} in Fig.~\ref{fig:air_snr} for an ISI-free channel with independent CSCG inputs. We also see that superposed pilots outperform interleaved pilots. The curves $C_\text{coh}$ show the coherent channel capacity, i.e., the \gls{air} with independent \gls{cscg} inputs.

\begin{figure}[t!]
    \centering
    \scalebox{\figscale}{
    \begin{tikzpicture}
    \begin{axis}[%
    width=7cm,
    height=4.5cm,
    at={(0,0)},
    scale only axis,
    xmin=-20,
    xmax=0,
    xlabel={$10\log_{10}\ratiopilots$ [dB]},
    ymin=0,
    ymax=3.5,
    ylabel={AIR [bpcu]},
    axis x line*=bottom,
    axis y line*=left,
    xmajorgrids,
    ymajorgrids,
    legend style={at={(0,0)},anchor=south west,nodes={scale=0.7, transform shape}},
    legend columns = 2,
    ]                

        \addplot [color=mycolor1, mark = |]
          table[]{figures/data/13dB_5e-3_air_v_k_cd_uniform_interleaved_16.tsv};
            \addlegendentry{SPA, interleaved};
            
        \addplot [color=mycolor1, dashed, mark = |, forget plot]
          table[]{figures/data/13dB_5e-3_air_v_k_cd_uniform_interleaved_64.tsv};
            
        \addplot [color=mycolor3, mark = x]
          table[]{figures/data/13dB_5e-3_air_v_k_cd_uniform_offset_16.tsv};
            \addlegendentry{SPA, superposed};
            
        \addplot [color=mycolor3, dashed, mark = x, forget plot]
          table[]{figures/data/13dB_5e-3_air_v_k_cd_uniform_offset_64.tsv};

        \addplot [color=mycolor4, mark = square]
              table[]{figures/data/13dB_5e-3_air_v_k_cd_uniform_LMMSE_L_12_16.tsv};
            \addlegendentry{LMMSE, 25};

        \addplot [color=mycolor4, dashed, mark = square, forget plot]
              table[]{figures/data/13dB_5e-3_air_v_k_cd_uniform_LMMSE_L_12_64.tsv};

        \addplot [color=mycolor5, mark = square]
              table[]{figures/data/13dB_5e-3_air_v_k_cd_uniform_LMMSE_16.tsv};
            \addlegendentry{LMMSE, $\infty$};

        \addplot [color=mycolor5, dashed, mark = o, forget plot]
              table[]{figures/data/13dB_5e-3_air_v_k_cd_uniform_LMMSE_64.tsv};

        \addplot [color=mycolor2, mark = *]
          table[]{figures/data/13dB_5e-3_air_v_k_cscg.tsv};
            \addlegendentry{SPA, no ISI};

        \addplot [color=black,thick,dashed]
          table[]{figures/data/13dB_coherent_capacity_v_k_cd.tsv};
        \addlegendentry{$C_\mathrm{coh}$};
        
    \end{axis}
\end{tikzpicture}%
    }
    \caption{Rates for \gls{ssmf} with SNR \SI{13}{\dB}, $\nu_\Delta=5\cdot 10^{-3}$, and 16-\gls{qam} (solid) and 64-\gls{qam} (dashed).}
    \label{fig:AIR_v_k_CD_uniform}
\end{figure}
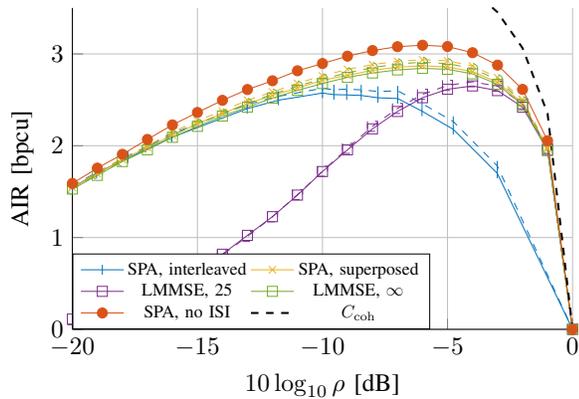
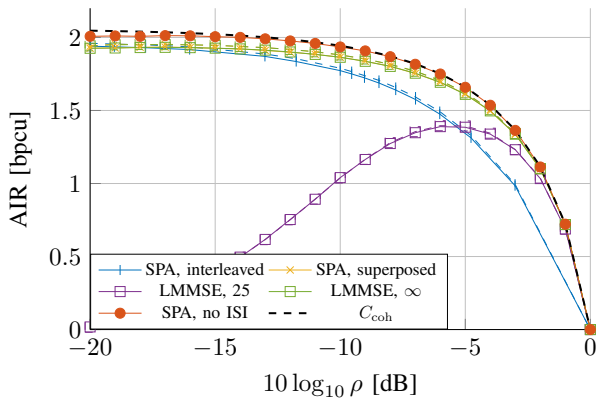
\begin{figure}[t!]
    \centering
    \scalebox{\figscale}{
    \begin{tikzpicture}
    \begin{axis}[%
    width=7cm,
    height=4.5cm,
    at={(0,0)},
    scale only axis,
    xmin=-20,
    xmax=0,
    xlabel={$10\log_{10}\ratiopilots$ [dB]},
    ymin=0,
    ymax=2.2,
    ylabel={AIR [bpcu]},
    axis x line*=bottom,
    axis y line*=left,
    xmajorgrids,
    ymajorgrids,
    legend style={at={(0,0)},anchor=south west,nodes={scale=0.7, transform shape}},
    legend columns = 2,
    ]                
        
        \addplot [color=mycolor1, mark = |]
          table[]{figures/data/5dB_1e-6_air_v_k_cd_uniform_interleaved_16.tsv};
            \addlegendentry{SPA, interleaved};
            
        \addplot [color=mycolor1, dashed, mark = |, forget plot]
          table[]{figures/data/5dB_1e-6_air_v_k_cd_uniform_interleaved_64.tsv};
            
        \addplot [color=mycolor3, mark = x]
          table[]{figures/data/5dB_1e-6_air_v_k_cd_uniform_offset_16.tsv};
            \addlegendentry{SPA, superposed};
            
        \addplot [color=mycolor3, dashed, mark = x, forget plot]
          table[]{figures/data/5dB_1e-6_air_v_k_cd_uniform_offset_64.tsv};

        \addplot [color=mycolor4, mark = square]
              table[]{figures/data/5dB_1e-6_air_v_k_cd_uniform_LMMSE_L_12_16.tsv};
            \addlegendentry{LMMSE, 25};

        \addplot [color=mycolor4, dashed, mark = square, forget plot]
              table[]{figures/data/5dB_1e-6_air_v_k_cd_uniform_LMMSE_L_12_64.tsv};

        \addplot [color=mycolor5, mark = square]
              table[]{figures/data/5dB_1e-6_air_v_k_cd_uniform_LMMSE_16.tsv};
            \addlegendentry{LMMSE, $\infty$};

        \addplot [color=mycolor5, dashed, mark = square, forget plot]
              table[]{figures/data/5dB_1e-6_air_v_k_cd_uniform_LMMSE_64.tsv};

            \addplot [color=mycolor2, mark = *]
          table[]{figures/data/5dB_1e-6_air_v_k_cscg.tsv};
            \addlegendentry{SPA, no ISI};

        \addplot [color=black,thick,dashed]
          table[]{figures/data/5dB_coherent_capacity_v_k_cd.tsv};
        \addlegendentry{$C_\mathrm{coh}$};
        
    \end{axis}
\end{tikzpicture}%
     }
    \caption{Rates for \gls{ssmf} with SNR \SI{5}{\dB}, $\nu_\Delta=10^{-6}$, and 16-\gls{qam} (solid) and 64-\gls{qam} (dashed).}
    \label{fig:AIR_v_k_CD_shaped}
\end{figure}

\subsection{Multipath Channel with \gls{ofdm}}
\label{subsec:multi-path-channel}
Consider the Proakis-C channel \cite[p.\ 654]{Proakis-Salehi-08} with taps
\begin{align}
   \v h_c^\transpose = [0.227, 0.460, 0.688, 0.460, 0.227].
   \label{eq:proakis-c}
\end{align}
The channel matrix $\m H_c$ in \eqref{eq:ofdm_channel} is circulant with rows $\v h_c^\transpose$ and we applied the \gls{dft} in \eqref{eq:linear-equalizer} to $\v Y'$. We use a single-tone pilot placed where the channel gain is largest, i.e., tone zero for the taps \eqref{eq:proakis-c}. This frequency bin is not used for messaging. The tone corresponds to a superposed pilot in the time domain.

Suppose the transmitter knows $\v h_c$ and performs waterfilling power allocation across the tones, which achieves capacity for \gls{cscg} symbols. We remark that the \gls{qam} performance can be improved by using other power allocations \cite{Zogakis-COMM95,Lozano-ISIT05}. \figref{fig:AIR_v_k_ofdm_13} plots the \glspl{air} for uniform \gls{qam} against the \gls{psr}. The variance \eqref{eq:extrinsic} is inaccurate, because the surrogate prior $q(\v z)$ uses independent $T_i$, which is inaccurate for \gls{ofdm} where $\v T$ has correlated entries. We computed it numerically.

\begin{figure}[t!]
    \centering
    \scalebox{\figscale}{
    \begin{tikzpicture}
    \begin{axis}[%
    width=7cm,
    height=4.5cm,
    at={(0,0)},
    scale only axis,
    xmin=-20,
    xmax=0,
    xlabel={$10\log_{10}\ratiopilots$ [dB]},
    ymin=0,
    ymax=1.8,
    ylabel={AIR [bpcu]},
    axis x line*=bottom,
    axis y line*=left,
    xmajorgrids,
    ymajorgrids,
    legend style={at={(0,0)},anchor=south west,nodes={scale=0.7, transform shape}},
    legend columns = 2,
    ]

        \addplot [color=mycolor1, mark = |]
          table[]{figures/data/13dB_5e-3_air_v_k_ofdm_waterfilling_uniform_sinusoid_16.tsv};
        \addlegendentry{SPA, tone zero};

        \addplot [color=mycolor1, dashed, mark = |, forget plot]
          table[]{figures/data/13dB_5e-3_air_v_k_ofdm_waterfilling_uniform_sinusoid_64.tsv};

        \addplot [color=mycolor4, mark = square]
          table[]{figures/data/13dB_5e-3_air_v_k_ofdm_waterfilling_uniform_LMMSE_L_12_16.tsv};
        \addlegendentry{LMMSE, 25};

        \addplot [color=mycolor4, dashed, mark = square, forget plot]
          table[]{figures/data/13dB_5e-3_air_v_k_ofdm_waterfilling_uniform_LMMSE_L_12_64.tsv};

        \addplot [color=mycolor5, mark = square]
          table[]{figures/data/13dB_5e-3_air_v_k_ofdm_waterfilling_uniform_LMMSE_16.tsv};
        \addlegendentry{LMMSE, $\infty$};

        \addplot [color=mycolor5, dashed, mark = square, forget plot]
          table[]{figures/data/13dB_5e-3_air_v_k_ofdm_waterfilling_uniform_LMMSE_64.tsv};

        \addplot [color=black,thick,dashed]
          table[]{figures/data/13dB_coherent_capacity_v_k_ofdm_waterfilling.tsv};
        \addlegendentry{$C_\mathrm{coh}$};
                
    \end{axis}
\end{tikzpicture}%
    }
    \caption{Rates for a multipath channel with SNR \SI{13}{\dB}, $\nu_\Delta=5\cdot 10^{-3}$, and 16-\gls{qam} (solid) and 64-\gls{qam} (dashed).}
    \label{fig:AIR_v_k_ofdm_13}
\end{figure}
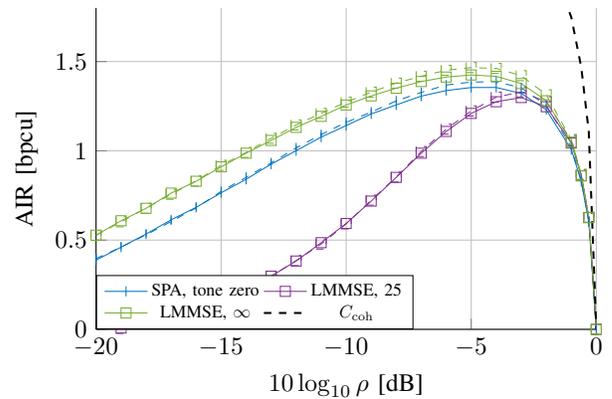
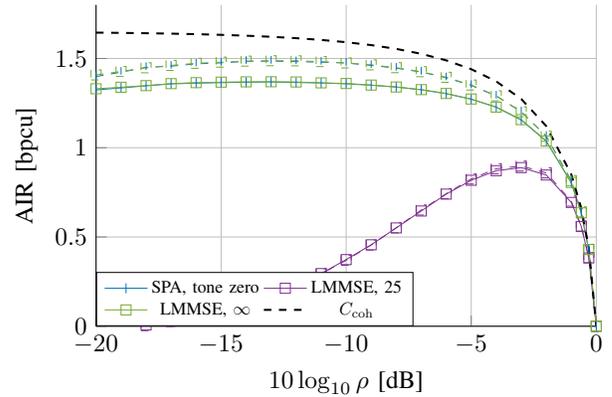
\begin{figure}[t!]
    \centering
    \scalebox{\figscale}{
    \begin{tikzpicture}
    \begin{axis}[%
    width=7cm,
    height=4.5cm,
    at={(0,0)},
    scale only axis,
    xmin=-20,
    xmax=0,
    xlabel={$10\log_{10}\ratiopilots$ [dB]},
    ymin=0,
    ymax=1.8,
    ylabel={AIR [bpcu]},
    axis x line*=bottom,
    axis y line*=left,
    xmajorgrids,
    ymajorgrids,
    legend style={at={(0,0)},anchor=south west,nodes={scale=0.7, transform shape}},
    legend columns = 2,
    ]

        \addplot [color=mycolor1, mark = |]
          table[]{figures/data/5dB_1e-6_air_v_k_ofdm_waterfilling_uniform_sinusoid_16.tsv};
        \addlegendentry{SPA, tone zero};

        \addplot [color=mycolor1, dashed, mark = |, forget plot]
          table[]{figures/data/5dB_1e-6_air_v_k_ofdm_waterfilling_uniform_sinusoid_64.tsv};

        \addplot [color=mycolor4, mark = square]
          table[]{figures/data/5dB_1e-6_air_v_k_ofdm_waterfilling_uniform_LMMSE_L_12_16.tsv};
        \addlegendentry{LMMSE, 25};

        \addplot [color=mycolor4, dashed, mark = square, forget plot]
          table[]{figures/data/5dB_1e-6_air_v_k_ofdm_waterfilling_uniform_LMMSE_L_12_64.tsv};

        \addplot [color=mycolor5, mark = square]
          table[]{figures/data/5dB_1e-6_air_v_k_ofdm_waterfilling_uniform_LMMSE_16.tsv};
        \addlegendentry{LMMSE, $\infty$};

        \addplot [color=mycolor5, dashed, mark = square, forget plot]
          table[]{figures/data/5dB_1e-6_air_v_k_ofdm_waterfilling_uniform_LMMSE_64.tsv};
        
        \addplot [color=black,thick,dashed]
          table[]{figures/data/5dB_coherent_capacity_v_k_ofdm_waterfilling.tsv};
        \addlegendentry{$C_\mathrm{coh}$};

    \end{axis}
\end{tikzpicture}%
    }
    \caption{Rates for a multipath channel with SNR \SI{5}{\dB}, $\nu_\Delta=10^{-6}$, and 16-\gls{qam} (solid) and 64-\gls{qam} (dashed).}
    \label{fig:AIR_v_k_ofdm_5}
\end{figure}

Note that we neglected the cyclic prefix. One can place pilots outside the \gls{dft} window, i.e., in the ``ramp up'' and ``ramp down'' phases of the linear channel convolution. Such symbols would improve the \gls{pn} compensation. 

\section{Conclusions}
\label{sec:conclusions}
We applied a non-iterative \gls{pn} compensation algorithm to \gls{isi} channels with Wiener \gls{pn}. The algorithm outperforms \gls{lmmse} filters of similar complexity and performs close to an infinite-tap filter. Superposed pilots in the time domain outperform interleaved pilots.

There are many directions for future work. For example, one can use multi-level coding and \gls{sic} as in \cite{Guess:00:ISI,Pfister:01,prinz2023successive,Jaeger-JSAC25} or iterations between the \gls{pn} and \gls{isi} compensation modules. However, both methods introduce latency. Next, one could use correlated priors and include the cyclic prefix in the factor graph to improve the \gls{ofdm} \gls{pn} compensation. Finally, for large phase noise, one should study more accurate channel models based on filtered continuous-time \gls{pn}; see e.g. 
\cite{Foschini1988Comm,Foschini1988IT,FoschiniCom1989} and \cite{LucaWhiteISIT2014,BarlettaISIT2015,Ghozlan-IT17}.

\section*{Acknowledgment}
This work was supported by the German Federal Ministry of Education and Research (BMBF) under project 16KISK002 and by the German Research Foundation (DFG) under Project KR 3517/13-1.

\clearpage
\bibliographystyle{IEEEtran}
\bibliography{bibliograpy}

\end{document}